\documentstyle[prl,aps]{revtex}
\input BoxedEPS.tex
\SetOzTeXEPSFSpecial
\HideDisplacementBoxes

\begin{document}

\twocolumn[\hsize\textwidth\columnwidth\hsize
           \csname @twocolumnfalse\endcsname
\title{Evidence for small or intermediate-size polarons in
the ferromagnetic state of manganites}
\author{Guo-meng  Zhao$^{(1,2)}$, D. J. Kang$^{(1)}$, W. Prellier$^{(1,*)}$,
M. Rajeswari$^{(1)}$, }
\author{H. Keller $^{(2)}$, T. Venkatesan $^{(1)}$ and  R. L. Greene$^{(1)}$}
\vspace{1cm}
\address{$^{(1)}$Center for Superconductivity Research, University of
Maryland, College Park, MD 20742,
USA\\
$^{(2)}$ Physik-Institut der Universit\"at Z\"urich, CH-8057
Z\"urich, Switzerland}

\maketitle
\noindent
\begin{abstract}
Oxygen-isotope effects on the intrinsic resistivity have been
studied in high-quality epitaxial thin films of
La$_{0.75}$Ca$_{0.25}$MnO$_{3}$ and Nd$_{0.7}$Sr$_{0.3}$MnO$_{3}$. We
found that the residual resistivity $\rho_{o}$ increases
by about 15(3)$\%$ upon replacing $^{16}$O by $^{18}$O. This provides
strong evidence for the presence of small or
intermediate-size polarons in the
metallic ferromagnetic state. Furthermore, the temperature dependent
part of the resistivity at low temperatures consists of an  $AT^{4.5}$
term contributed from
2-magnon scattering, and a
$B\omega_{s}/\sinh^{2}(\hbar\omega_{s}/2k_{B}T)$ term which arises from
scattering by a soft optical phonon mode. The absolute magnitudes of
the coefficient $A$ and the phonon frequency $\hbar\omega_{s}$ for both
isotope samples are in
quantitative agreement with theoretical predictions.
\end{abstract}
\vspace{1cm}
]

\narrowtext
The magnetic properties of the manganite
perovskites Re$_{1-x}$A$_{x}$MnO$_{3}$ (Re = a rare-earth element, and A =
a divalent
element) have attracted renewed
interest
because of the observation of colossal
magnetoresistance (CMR) in thin films of these materials
\cite{Von}. Despite tremendous experimental efforts \cite{Art},
the basic physics and the microscopic mechanism for the colossal
magnetoresistance in these materials remain
controversial \cite{Millis1,Moreo,Alex}. In particular, the
nature of the charge carriers and
the electrical transport mechanism in the low-temperature metallic
state have not been fully understood.

At low temperatures, a dominant
$T^{2}$ contribution in resistivity is generally observed, and has
been ascribed to electron-electron scattering \cite{Urushibara}.
In contrast, Jaime {\em et al.} \cite{Jaime2} have
shown that the resistivity is essentially temperature
independent below 20 K and exhibits a strong $T^{2}$ dependence above
50 K. They proposed single magnon scattering with a cutoff at
long wavelengths to explain their data. In their scenario \cite{Jaime2}, they
considered a
case where the manganese $e_{g}$ minority (spin-up) band lies slightly above
the Fermi level (in the majority spin-down band)  with a
small energy gap of about 1 meV.  This is in contradiction with the optical
data
\cite{Machida} which show that the manganese $e_{g}$ minority band is well
above the Fermi level. Alternatively, Zhao {\em et al.}
\cite{ZhaoPRL00} have recently shown that the temperature dependent
part of the resistivity at low temperatures
is mainly due to scattering by a soft optical phonon mode.

Various CMR theories \cite{Millis1,Moreo,Alex} predict different natures of
charge carriers
in the ferromagnetic
state, namely, large polarons vs small polarons.
The clarification of the nature of
charge carriers in the
ferromagnetic state can discriminate those theoretical models. One way
to address this issue is to study the isotope effect on the effective
carrier mass $m^{*}$. A strong isotope dependence of $m^{*}$
indicates the presence of small polaronic carriers. This is because
the effective mass of small polaronic carriers is enhanced by a
factor $\exp (\Gamma E_{p}/\hbar\omega_{o})$, which depends strongly on
the isotope mass $M$ if $\Gamma E_{p}/\hbar\omega_{o}$ is substantial
(where $E_{p}$ is the polaron binding energy independent of $M$, and
the characteristic optical phonon frequency $\omega_{o}$ $\propto$
1/$\sqrt{M}$) \cite{ZhaoNature,Alex99}.

Here  we report studies of the oxygen-isotope effect
on the intrinsic low-temperature resistivity in high-quality
epitaxial thin films of
La$_{0.75}$Ca$_{0.25}$MnO$_{3}$ and
Nd$_{0.7}$Sr$_{0.3}$MnO$_{3}$. The residual resistivity of
these compounds shows a strong dependence on the oxygen-isotope mass,
which provides clear evidence for the presence of small or
intermediate-size polarons in the ferromagnetic state.


Epitaxial thin films of La$_{0.75}$Ca$_{0.25}$MnO$_{3}$ (LCMO) and
Nd$_{0.7}$Sr$_{0.3}$MnO$_{3}$ (NSMO) were grown
on $<$100$>$ LaAlO$_{3}$ single crystal substrates by pulsed laser
deposition using a KrF excimer laser \cite{Prellier}.
The film thickness was about 190 nm for NSMO and 150 nm for LCMO.
Two halves were cut
from the same piece of a film for oxygen-isotope diffusion.
The diffusion for LCMO/NSMO was
carried out for 10 h
at about 940/900 $^{\circ}$C and oxygen pressure of 1 bar. The
$^{18}$O-isotope gas is enriched with
95$\%$ $^{18}$O, which can ensure 95$\%$ $^{18}$O in the $^{18}$O thin
films. The resistivity was measured using the van der Pauw technique, and the
contacts were made by silver paste. The measurements were
carried out in a Quantum Design measuring system.

Fig.~1 shows the resistivity of the oxygen-isotope exchanged films of
La$_{0.75}$Ca$_{0.25}$MnO$_{3}$ (LCMO) and
Nd$_{0.7}$Sr$_{0.3}$MnO$_{3}$ (NSMO) over 100-300 K. It
is apparent that the $^{18}$O samples have
lower metal-insulator crossover temperatures and much sharper
resistivity drop. The Curie temperature $T_{C}$ normally coincides with a
temperature where $d\ln\rho/dT$ exhibits a maximum. We find that the
oxygen-isotope shift of $T_{C}$ is 14.0(6) K for LCMO, and 17.5(6)
K for NSMO, in excellent agreement with the
results for the bulk samples \cite{Zhao99}.
\begin{figure}[htb]
    \ForceWidth{6.6cm}
	\centerline{\BoxedEPSF{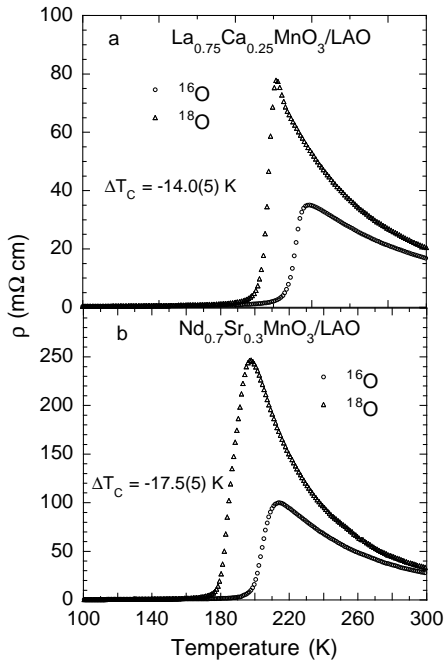}}
	\vspace{0.3cm}
	\caption[~]{The resistivity of the oxygen-isotope exchanged
	films of (a) La$_{0.75}$Ca$_{0.25}$MnO$_{3}$; (b)
Nd$_{0.7}$Sr$_{0.3}$MnO$_{3}$.}
	\protect\label{Fig.1}
\end{figure}

In Fig.~2 we plot the low-temperature resistivity of the oxygen-isotope
exchanged films of (a) LCMO; (b) NSMO. In both cases, the
residual resistivity $\rho_{o}$ for the $^{18}$O samples is larger than
for the $^{16}$O samples by about 15$\%$. We have repeated the van
der Pauw measurements at 5 K several times with different
contact configurations. We checked that the uncertainty of the
difference in $\rho_{o}$ of the two isotope samples is
less than 3$\%$. We also found that the isotope effect is reversible
upon isotope back-exchange. It is worth noting that the $\rho_{o}$
for our LCMO $^{16}$O film is similar to that for a single crystal
with a similar $T_{C}$ \cite{Dai}, while the $\rho_{o}$
for our NSMO $^{16}$O film is about 40$\%$ higher
than those for single crystals \cite{Dai,Sawaki}. This indicates that the
electrical transport observed in our LCMO films is intrinsic. On the
other hand, a larger $\rho_{o}$ and a
small upturn in the low-temperatrure resistivity of our NSMO films are
most likely to arise from carrier localization when the
low-temperature resistivity is larger than a critical value of about
300 $\mu\Omega$cm  \cite{Sawaki}. The larger $\rho_{o}$ in our NSMO
films might be caused by a contamination of impurities (e.g., Al)
that may diffuse from the LAO substrate during the high-temperature
annealing. We should mention that
the intrinsic resistivity cannot be obtained from ceramic samples
where the boundary resistivity is dominant. Thus
one cannot use ceramic samples to study the isotope effect on the
intrinsic resistivity. Moreover, the van der Pauw technique is
particularly good to precisely measure the resistivity difference between the
oxygen-isotope exchanged films which have the same thickness. Thus
the data shown in Fig.~2 represent the first precise measurements on
the intrinsic resistivity of the isotope substituted samples.
\begin{figure}[htb]
\ForceWidth{6.6cm}
	\centerline{\BoxedEPSF{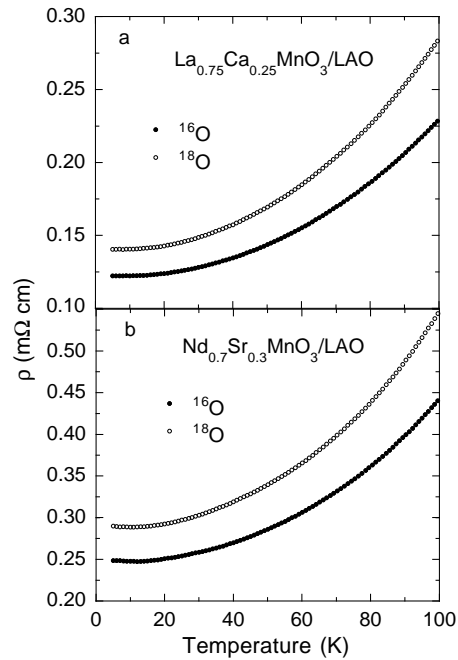}}
	\vspace{0.3cm}
	\caption[~]{The low-temperature resistivity of the oxygen-isotope
exchanged
	films of (a) La$_{0.75}$Ca$_{0.25}$MnO$_{3}$; (b)
Nd$_{0.7}$Sr$_{0.3}$MnO$_{3}$.}
	\protect\label{Fig.2}
\end{figure}

It was shown that \cite{Pick} the intrinsic residual resistivity
$\rho_{o}$ $\propto$ $m^{*}/n\tau_{o}$, where $\hbar/\tau_{o}$ is
the scattering rate which is associated with the random potential
produced by randomly distributed trivalent and divalent cations, $m^{*}$ is
the effective mass of carriers at low temperatures, and $n$ is the mobile
carrier
concentration. If
$\hbar/\tau_{o}$ is nearly independent of $m^{*}/n$,
$\rho_{o}$ should be proportional to $m^{*}/n$, or to the electronic
specific heat coefficient
$\gamma$ for a fixed $n$. Comparing the intrinsic $\rho_{o}$ values in
the best single crystals of Nd$_{0.7}$Sr$_{0.3}$MnO$_{3}$ and
La$_{0.7}$Sr$_{0.3}$MnO$_{3}$ with their $\gamma$ values ($\rho_{o}$
$\simeq$ 170 $\mu\Omega$cm
\cite{Dai,Sawaki} and $\gamma$ = 22(3) mJ/moleK$^{2}$ for
Nd$_{0.7}$Sr$_{0.3}$MnO$_{3}$ \cite{Gordon}; $\rho_{o}$ $\simeq$
30$\mu\Omega$cm \cite{Sawaki} and $\gamma$ = 3.5(5) mJ/moleK$^{2}$ for
La$_{0.7}$Sr$_{0.3}$MnO$_{3}$ \cite{Okuda}),  we find that
$\rho_{o}$ is nearly proportional to $\gamma$ or $m^{*}$. This
suggests that the
observed large oxygen-isotope effect on $\rho_{o}$ is mainly caused by the
isotope
dependence of $m^{*}$, rather than of the scattering rate.

If the charge carriers at low temperatures are of
small or intermediate-size polarons, the temperature dependence of the
resistivity should agree with polaron metallic conduction. This
is indeed the case as recently demonstrated by Zhao {\em et al.}
\cite{ZhaoPRL00}. There are three contributions to the
resistivity: the residual resistivity
$\rho_{o}$, the term $AT^{4.5}$ contributed from 2-magnon scattering
\cite{Kubo}, and the term
$B\omega_{s}/\sinh^{2}(\hbar\omega_{s}/2k_{B}T)$, which arises from
polaron coherent motion involving a relaxation due to a soft optical phonon
mode that is strongly coupled to the carriers \cite{ZhaoPRL00}. Here
$\omega_{s}$ is
the frequency of a soft optical
mode. The temperature dependent part of the
resistivity is then given by
\begin{equation}
\rho(T) - \rho_{o} = AT^{4.5}+ B\omega_{s}/\sinh^{2}(\hbar\omega_{s}/2k_{B}T).
\end{equation}

\begin{figure}[htb]
\ForceWidth{6.6cm}
	\centerline{\BoxedEPSF{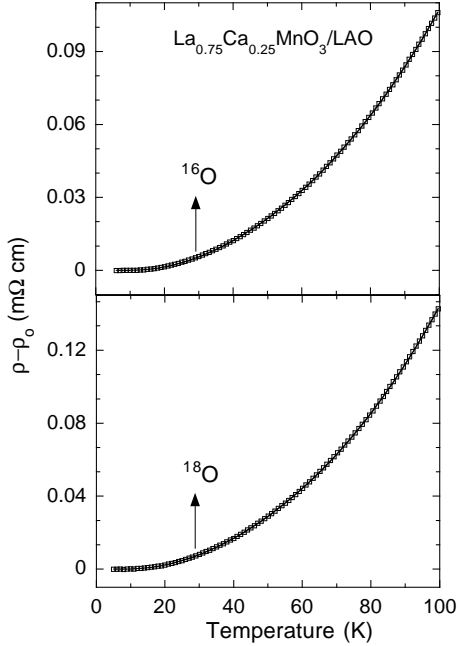}}
	\vspace{0.3cm}
	\caption[~]{$\rho(T) - \rho_{o}$ for the $^{16}$O and $^{18}$O films
	of La$_{0.75}$Ca$_{0.25}$MnO$_{3}$. The solid lines are fitted
	curves by Eq.~1.}
	\protect\label{Fig.3}
\end{figure}
It was shown that the parameter $B$ is
proportional to $m^{*}/n$ \cite{ZhaoPRL00}. This implies
that $B$ and $\rho_{o}$ should have the same relative change upon the
isotope substitution if the scattering rate $\hbar/\tau_{o}$ is
isotope independent. The coefficient $A$ has an
analytical expression
in the case of a simple parabolic conduction band (occupied by
single-spin holes) \cite{Kubo}. In terms of the hole density per cell $n$,
the average spin stiffness $D$, and the effective hopping integral
$t^{*}$,  the coefficient $A$ can be written as \cite{Kubo}
\begin{eqnarray}
&A =& (\frac{3a\hbar}{32\pi e^{2}})(2 - n/2)^{-2}(6\pi^{2}n)^{5/3}(2.52 +
0.0017\frac{D}{a^{2}t^{*}})
\nonumber \\
& &\{\frac{a^{2}k_{B}}{D(6\pi^{2})^{2/3}(0.5^{2/3}-n^{2/3})}\}^{9/2}.
\end{eqnarray}
Here we have used the relations: $ak_{F}= (6\pi^{2}n)^{1/3}$ (where
$\hbar k_{F}$ is the Fermi momentum, and $a$ is the lattice constant);
$E_{F}= t^{*}(6\pi^{2})^{2/3}(0.5^{2/3}-
n^{2/3})$ (where the Fermi energy $E_{F}$ is measured from the band
center); the effective spin $S^{*} = 2 - n/2$. The value of $t^{*}$ can be
estimated to be about 40 meV
from the measured effective plasma frequency $\hbar\Omega_{p}^{*}$ =
1.1 eV and $n \sim$ 0.3 in La$_{0.7}$Ca$_{0.3}$MnO$_{3}$ \cite{Simpson}.
Since $D$ is
about 100 meV \AA$^{2}$ (see below), one expects that the term
0.0017$D/a^{2}t^{*}$$<$$<$ 2.52, and thus can be dropped out in
Eq.~2. Then there are two parameters $n$ and
$D$ that determine the magnitude of $A$. In doped manganites, $n$
should be approximately equal to the doping level $x$.
The average spin stiffness $D$ should be close to the long-wave spin
stiffness $D(0)$ if there is
negligible magnon softening near the zone boundary.
On the other hand, $D$ $<$ $D(0)$ if
there is a magnon softening near the zone boundary as the case of
low $T_{C}$ materials \cite{Hwang}. In any cases, one might expect
that the average $D$ should be proportional to $T_{C}$ so that
$D/k_{B}T_{C}$ is a univeral constant in the manganite system. Since the
magnon
softening becomes unimportant when the $T_{C}$ is higher than 350 K
\cite{Hwang}, then $D$ $\simeq$ $D(0)$ in the
compound La$_{0.7}$Sr$_{0.3}$MnO$_{3}$ with the highest $T_{C}$ = 378
K \cite{Martin}. Thus, the universal value of $D/k_{B}T_{C}$ should be
close to the value of $D(0)/k_{B}T_{C}$ in
La$_{0.7}$Sr$_{0.3}$MnO$_{3}$, which was found
to 5.8$\pm$0.2 \AA$^{2}$ \cite{Martin}.
\begin{figure}[htb]
\ForceWidth{6.6cm}
	\centerline{\BoxedEPSF{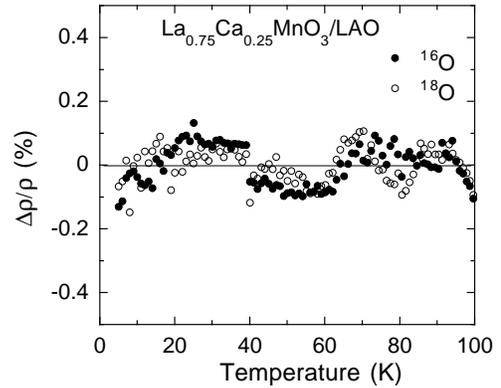}}
	\vspace{0.3cm}
	\caption[~]{The relative differences between the resistivity data and
	the fitted curves for the $^{16}$O and $^{18}$O films
	of La$_{0.75}$Ca$_{0.25}$MnO$_{3}$. The systematic
	deviations are very small.}
	\protect\label{Fig.4}
\end{figure}

Now we fit the $\rho(T) - \rho_{o}$ data below 100 K by Eq.~1 for the
LCMO $^{16}$O and $^{18}$O samples, as shown in Fig.~3. It is striking
that the fits to the data of both isotope samples are very good. In order to
see more clearly the quality of the fits, we plot in Fig.~4 the relative
difference $\Delta \rho/\rho$
between the
data and the fitted curves. It is clear that there is a negligible
systematic deviation below 100 K. We exclude the data above 100 K in
the fitting since $n/m^{*}$ above 100 K becomes temperature
dependent \cite{Simpson}.
\begin{table}[htb]
	\caption[~]{The summary of the fitting parameters $A$, $B$ and
	$\hbar\omega_{s}/k_{B}$ for the
	$^{16}$O and $^{18}$O films of La$_{0.75}$Ca$_{0.25}$MnO$_{3}$
	(LCMO), and the summary of the $T_{C}$ and $\rho_{o}$ values for
the LCMO
	films and the Nd$_{0.7}$Sr$_{0.3}$MnO$_{3}$ films (NSMO).
	The uncertainty in $T_{C}$ is $\pm$0.3 K. The uncertainty in
	$\rho_{o}$ is discussed in the text.}
	\begin{center}
    \begin{tabular}{lccccc}
    	Compounds &$T_{C}$ & $\rho_{0}$ &$A$ & $B$
    	&$\hbar\omega_{s}/k_{B}$
    	 \\
    	 &(K) &($\mu\Omega$cm) &(m$\Omega$cm/K$^{4.5}$)&($\mu\Omega$cm/K) &
    	 (K)\\
    	\hline
    	LCMO($^{16}$O)&231.3 &122.4 &1.20(2)$\times$10$^{-11}$ & 0.370(3)&
    	74.4(2)\\
    	LCMO($^{18}$O)&217.3 &140.5 &1.89(2)$\times$10$^{-11}$ &
    	0.434(3)&66.8(3)\\
    NSMO($^{16}$O)&203.9 &248.2& & & \\
    NSMO($^{18}$O)&186.4 &289.2 & & & \\
\end{tabular}
\end{center}
\protect\label{Tab1}
\end{table}
\noindent
The fitting
parameters $A$, $B$ and $\hbar\omega_{s}$ are summarized in Table I. Since
the fits are excellent, the uncertainties in the fitting parameters are
very small (see Table I).  On the other hand,
since there is
a small upturn at the low-temperature resistivity of the NSMO films
(see Fig.~2 and discussion above), one needs three additional parameters
in order to fit the low temperature upturn. Therefore, we did not attempt to
fit the NSMO
data with six parameters, but gave the values of the
resistivity at 5 K for both isotope samples in Table I.

From Table I, one can see that $\rho_{o}$ increases by 15(3)$\%$, and
$B$ by 17(3)$\%$. This provides additional evidence that the scattering
rate $\hbar/\tau_{o}$ is
nearly isotope-mass independent, in agreement with the above
argument. Thus the observed large oxygen-isotope effects on both
$\rho_{o}$ and $B$ suggest that the effetive mass of carriers depends
strongly on the oxygen-isotope mass. This is consistent with the
presence of small or intermediate-size polarons in the ferromagnetic
state of manganites.

In addition, $\omega_{s}$ decreases by about 10(1)$\%$
upon replacing $^{16}$O with $^{18}$O. This may imply that the soft
mode might be associated with the motion of the oxygen atoms and has
a large anharmonicity. It was shown that the tilt/rotation mode of the
oxygen octahedra in cuprates has a strong electron-phonon coupling (quadratic
coupling) and a large anharmonicity \cite{Vincent,Pickett}. The large
anharmonicity of the mode can possibly lead to a decrease of $\omega_{s}$ by
12.5$\%$ upon replacing $^{16}$O with $^{18}$O \cite{Vincent}. In a similar
perovskite superconductor Ba(Pb$_{0.75}$Bi$_{0.25}$)O$_{3}$, both neutron and
tunneling experiments \cite{Reichardt} show that a soft mode with
$\hbar\omega_{s}/k_{B}$ = 70 K is related to rotational vibrations of
the oxygen
octahedra, and has a strong electron-phonon coupling. Moreover, the frequency
of the
rotational mode in Ba(Pb$_{0.75}$Bi$_{0.25}$)O$_{3}$ is nearly the
same as that of the soft mode
($\hbar\omega_{s}/k_{B}$ = 74 K) in
the La$_{0.75}$Ca$_{0.25}$MnO$_{3}$ $^{16}$O
film. Both the frequency of the soft mode and its isotope dependence
can be quantitatively explained if the soft mode in the manganites is also
associated
with the rotational vibrations of
the oxygen octahedra.

Now we turn to the discussion on the magnitude of the parameter $A$
and its isotope dependence. From Eq.~2, one can see that $n$ and/or
$D$ should be isotope dependent in order to explain the large isotope
effect on the parameter $A$. As discussed above, the $D$ in Eq.~2 should be
proportional to $T_{C}$. Then one can
obtain $D$ values from the $T_{C}$ values, and the value of
$D/k_{B}T_{C}$ = 5.8 $\AA^{2}$ (see the above discussion). Substituting the
$D$ and $A$ values (see Table I) into Eq.~2, we find $n$ = 0.235 for the
La$_{0.75}$Ca$_{0.25}$MnO$_{3}$ $^{16}$O film, and $n$ =
0.240 for the La$_{0.75}$Ca$_{0.25}$MnO$_{3}$ $^{18}$O film.  The fact that
$n$ $\simeq$ $x$ for both isotope samples is
consistent with our previous interpretation of the isotope dependence
of $\rho_{o}$ being caused only by $m^{*}$. The $AT^{4.5}$ term
in our resistivity data
is in quantitative agreement with the 2-magnon scattering theory
\cite{Kubo}.

In summary, the oxygen-isotope effects observed in high-quality
epitaxial thin films of
La$_{0.75}$Ca$_{0.25}$MnO$_{3}$ and Nd$_{0.7}$Sr$_{0.3}$MnO$_{3}$
strongly suggest that the charge
carriers in the ferromagnetic state are small or intermediate-size
polarons. Moreover, the temperature dependent part of the resistivity
at low temperatures is in quantitative agreement with a transport
mechanism where the resistivity is due to two-magnon scattering
\cite{Kubo}
and scattering from a soft optical phonon mode. The results will place
important constraints on various CMR theories.

{\bf Acknowlegement}: The work was supported by the
NSF MRSEC at the University of Maryland and Swiss National Science
Foundation.
~\\
~\\
$*$ Present address: Laboratoire CRISMAT-ISMRA,14050 CAEN Cedex, France.
\bibliographystyle{prsty}



\end{document}